\documentclass[doublecol]{epl2} 
\usepackage{amssymb}
\title{Patch-repetition correlation length in glassy systems}
\shorttitle{} %Insert here a short version of the title if it exceeds 70 characters

\author{Chiara Cammarota \and Giulio Biroli}
\institute{Institut de Physique Th\'eorique (IPhT), CEA, and CNRS URA 2306, F-91191 Gif-sur-Yvette,
France}
\pacs{64.70.P}{Glass Transition}
\pacs{05.20.-y}{Statistical Mechanics classical}

\abstract{We obtain the patch-repetition entropy $\Sigma$ within the Random First Order Transition theory (RFOT) and for the square plaquette system, a model related to the dynamical facilitation theory of glassy dynamics. We find that in both cases the entropy of patches of linear size $\ell$, $\Sigma(\ell)$, 
scales as $s_c \ell^d+A\ell^{d-1}$ down to length-scales of the order of one, where $A$ is a positive constant, 
 $s_c$ is the configurational entropy density and $d$ the spatial dimension. In consequence, the only meaningful length that can be defined from patch-repetition is the cross-over length $\xi=A/s_c$. We relate $\xi$ to the typical length-scales already discussed in the literature and show that it is always of the order of the largest static length. 
Our results provide new insights, which are 
particularly relevant for RFOT theory, on the possible real space structure of super-cooled liquids. 
They suggest that this structure differs from a mosaic of different patches having roughly the same size.}

\begin{document}

\maketitle
The search for static amorphous order  and its characterization in glass-forming liquids has become 
a very active research topic in the last few years. 
Indeed, after that the understanding and the measurement of dynamical heterogeneity and dynamical 
correlation lengths reached a mature stage \cite{OUP}, the focus partially shifted to unveiling, measuring
and explaining static correlation lengths. This is a new and very promising way to understand the 
glass transition and prune down several theoretical explanations of the glassy behavior of super-cooled liquids.
In a nutshell, one would like to understand whether (and why) super-cooled liquids display increasing amorphous order 
and to what extent this phenomenon is related to the very fast growth of the relaxation time.\\
Many different static lengths have been proposed in the literature.
Here we only focus on the ones 
aimed at measuring growing static order---whatever this order is. This restricts the lengths available to the 
point-to-set \cite{BB,MS} and the patch-repetition \cite{KL} ones. Several numerical and analytical studies of the former have been already performed. Instead, very little is known on the latter,  
apart from what discussed in the original papers of Levine and Kurchan \cite{KL,KL2} and in 
a numerical work on a monodisperse Lennard-Jones liquid on the hyperbolic plane \cite{SL}. \\
Here we shall analytically study patch-repetition  within two frameworks: the Random First Order Transition (RFOT) theory \cite{RFOTreviewWolynes, RFOTreviewBB} and the square plaquette system, a model that is related to the dynamical facilitation theory 
of glassy dynamics \cite{JPG}. In \cite{KL,KL2} it was proposed that several lengths are encoded in the 
$\ell$ dependent behavior of the entropy, $\Sigma(\ell)$, measuring the repetition of patterns of linear size $\ell$: a cross-over length $\xi$ determining the regime 
in which $\Sigma(\ell)$ scales extensively, {\it i.e.} proportionally to $\ell^d$ ($d$ the spatial dimension) and a cooperativity length $\xi_c$ at which $\Sigma(\ell)$ becomes larger than one. It was suggested that a proxy for the  latter is provided by the configurational entropy $s_c$, which is the intensive value of the PR-entropy, to the power $-1/d$. 
In this work we find that within both frameworks
$\Sigma(\ell)$ scales as $s_c \ell^d+A\ell^{d-1}$ down to length-scales $\ell$ of the order of one, where $A$ is a positive constant. Within RFOT we assumed a surface tension exponent $\theta=d-1$, were this not the case 
the second term should be replaced by $A\ell^\theta$. 
The main consequence of the scaling we find for  $\Sigma(\ell)$ is that the only meaningful and growing static length that can be defined from patch-repetition is the cross-over length: $\xi=A/s_c$. By comparing $\xi$ to the typical length-scales already discussed in the literature, we find that $\xi$ always coincides with the largest static correlation length of the problem---a result possibly valid 
in general. The point-to-set length and $\xi$ are only equal in the case of RFOT. For the square plaquette model $\xi$ instead increases faster than the point-to-set length.
A final finding worth mentioning is that our results   
%Our result that $\Sigma(\ell)$ becomes of the order of one on a length-scale which remains microscopic and does not grow when decreasing the temperature is, at first sight, counter-intuitive. In particular, as we shall show later, it  
suggest that  that the real space structure of super-cooled liquids differs from a mosaic of different patches 
having roughly the same size, a result particular relevant for RFOT theory, as we shall discuss in detail at the end of 
this letter.  
\section{Patch-repetition entropy}
The patch repetition entropy was defined to probe amorphous order present in a given snapshot of a glassy liquid\footnote{Actually, also the point-to-set length could be in principle measured from a given snapshot but in reality this
is impractical; it is better to use ensemble averages.}. In order to obtain it, one has to count the frequency 
with which a given pattern repeats in a configuration. Actually, 
one should not count as different, patterns which only change because of short-time thermal fluctuations. 
This is a tricky issue that was solved in \cite{KL2}. For simplicity, we will neglect it below and discuss it 
later when needed.\\
The frequency with which a given pattern $\mathcal P$ repeats in a configuration $\mathcal C$ reads:
\begin{equation}
f_{\mathcal{P}}=\frac 1 V \int d{\mathbf r}\, \Omega({\mathcal C},{\mathcal P}_{\mathbf r}) \quad.
\end{equation}
where $V$ is the volume. The notation $\Omega({\mathcal C},{\mathcal P}_{\mathbf r})$ represents a delta function measuring whether the pattern $\mathcal P$ is present in $\mathcal C$ 
around the position $\mathbf r$. In the thermodynamic limit $f_{\mathcal{P}}$ coincides with its ensemble average since intensive quantities do not fluctuate. Thus, 
\begin{equation}
f_{\mathcal{P}}=\langle f_{\mathcal{P}} \rangle =P(\mathcal{P}) \quad,
\end{equation}
where $P(\mathcal{P})$ is the probability to observe the pattern $\mathcal P$ around a given point.
As usual for entropies, one prefers to focus on the logarithm of $P(\mathcal{P})$. Moreover, instead of 
focusing on each single pattern, a quite huge task, it is better to compute an average quantity that 
measures the repetition of {\it typical} patterns\footnote{Patterns identical by reflection or rotation will be consider as different for simplicity. This double counting leads to an error at most of the order 
of $\ln \ell$, which is irrelevant for our discussion.}. Following these recipes one ends up with the expression of the 
PR entropy of \cite{KL,SL}:
\begin{equation}
\Sigma(\ell)=-\sum_{\mathcal{P}}  P(\mathcal{P}) \log P(\mathcal{P})\quad.
\end{equation}
The sum over $\mathcal{P}$ is restricted to patterns appearing within a given region of linear size $\ell$. 
We shall focus on (hyper)-cubes or spheres, as done in \cite{SL}. 
This should not be an important restriction except if relevant patterns are very much diluted and fractal.\\
Levine and Kurchan proposed that at least two length-scales can be obtained in the super-cooled regime
by studying the dependence on $\ell$ of $\Sigma(\ell)$. Since amorphous order (if present) does not have an infinite range, one expects $\Sigma(\ell)$ to be extensive at large $\ell$, {\it i.e.} $\Sigma(\ell)\simeq s_c \ell^d$. Thus, 
one can define a cross-over length $\xi$ for which the extensive behavior, $\Sigma(\ell)/\ell^d\simeq s_c$, is attained. A second length $\xi_c$ (called cooperative in \cite{KL})  possibly different from the first one, can be defined as the largest length at which $\Sigma(\ell)$ is still of order one. The idea is that 
for a system characterized by infinite range amorphous order $\Sigma(\ell)$ never becomes extensive and remains of the order of one for any $\ell$ or just very slowly dependent on $\ell$, e.g. logarithmically. Thus, $\xi_c$ would be the largest scale over which $\Sigma(\ell)$ behaves as in the ideal 
glass phase; $\xi_c$ was thought to correspond to the spatial extent of medium range amorphous order. It was suggested \cite{KL2,SL} that an estimate of $\xi_c$ can be obtained by assuming that the extensive behavior is valid until $\Sigma(\ell)$ is of order one; this leads to $\xi_c=s_c^{-1/d}$. As we shall see, in the cases we have analyzed these assumptions
do not work. The only meaningful length is $\xi$.\\
Before concluding this introduction on PR lengths we would like to discuss another motivation to focus on $\xi$. 
In the works by Kurchan and Levine the study of pattern repetition was motivated by asking the question: {\it Suppose we have a region of volume V with a configuration A. To what extent does A determine the configuration (say B) of a neighboring region, also of size V?} This issue is more general than the one often addressed in analysis of 
point-to-set correlations, in which it is studied how much a set (a neighboring configuration, a boundary condition, etc)
determines the average density profile. It goes beyond that, because it asks to what extent $A$ determines $B$ 
without referring to any specific correlation function: for example $A$ could determine the three point correlation functions
in $B$ but not the average density profile. From this perspective, the answer to the question posed by Kurchan and Levine should lead to a static length always larger or equal to the point-to-set. In order to directly 
show that this length is actually $\xi$ let us rephrase the question in a more formal way: given the configuration $\mathcal{C}_A$ in $A$, how much the entropy of the configurations in $B$ is reduced? This can be obtained by subtracting to the entropy of $B$ the one obtained by constraining the configuration in $A$ to be equal to a typical 
configuration $\mathcal{C}_A$. This quantity is well known in information theory, it is called the mutual information and reads:
\begin{eqnarray}
I(A,B)&=&\Sigma(B)-\Sigma(B|A)=-\sum_{\mathcal{C}_B} P(\mathcal{C}_B) \log P(\mathcal{C}_B)\nonumber\\
&+&\sum_{\mathcal{C}_A}P(\mathcal{C}_A)\sum_{\mathcal{C}_B} P(\mathcal{C}_B|\mathcal{C}_A) \log P(\mathcal{C}_B|\mathcal{C}_A)
\end{eqnarray}
It is very easy to show that $I(A,B)=\Sigma(A)+\Sigma(B)-\Sigma(A+B)$. In the extreme case where $A$ determines $B$ completely, one obtains $\Sigma(B|A)=0$ and, hence, $\Sigma(A+B)=\Sigma(A)$. Instead, if $A$ does not determine $B$ at all, $\Sigma(B|A)=\Sigma(B)$ and $\Sigma(A+B)=\Sigma(A)+\Sigma(B)$. 
In consequence, in the ordered phase, one expects a PR-entropy that grows very slowly with $\ell$, whereas in the
super-cooled phase, which is characterized (at best) only by medium range amorphous order, the PR-entropy scales extensively
for large enough $\ell$. When $\ell$ reaches the value $\xi$ at which extensivity is attained the relative value of the mutual information, $[\Sigma(B)-\Sigma(B|A)]/\Sigma(B)$, vanishes. This means, recalling  Kurchan and Levine question, that $\xi$ can be interpreted as the length beyond which $A$ does not determine $B$ anymore. %FOr the conclusion: \footnote{The only caveat to this discussion, which also applies to all PR-lengths defined in the literature, is that the 
%length is obtained by studying the scaling of a global quantity and not a correlation function. Thus, below $\xi_a$ we know that the configuration in $B$ are influenced by the ones in $A$ but we do not have any spatial information on this dependence}.    
\section{Square Plaquette model}
This model consists in Ising spins interacting
with their neighbours through plaquette interactions. We consider 
the two dimensional plaquette model on the square lattice
\cite{JPG} whose Hamiltonian reads:
\begin{equation}
H= -J \sum_{ijkl \in \square} \sigma_i \, \sigma_j 
	\, \sigma_k \, \sigma_l
\label{H4}
\end{equation}
where the sum runs over all plaquettes of the lattice.  
The dynamics of this system can be shown to be effectively described by 
a kinetically constrained dynamics leading to an Arrhenius behavior. Its triangular 
counterpart is characterized by super-Arrhenius dynamics at low temperature \cite{JPG,Garrahan}.\\
By changing variables and defining $\tau_i=\sigma_i \, \sigma_j 
	\, \sigma_k \, \sigma_l$ the system becomes non-interacting. At low temperature almost all $\tau$s are up 
	and very few are down. The concentration of the latter is given by $c=(1-\tanh \beta J)/2\simeq e^{-2\beta J}$  
	The mapping 
	from $\sigma$s to $\tau$s is very non-linear. Thus, even though
	its thermodynamics is trivial, the model displays interesting features such as diverging static correlations at low temperature. The 2-point spin correlation function remains featureless but, instead, the 4 point correlation 
	function 
	\begin{equation}
	G_4(l)=\langle \sigma_{x,y}\sigma_{x,y+1}\sigma_{x+l,y}\sigma_{x+l,y+1}\rangle=\tanh (\beta J)^l\simeq e^{-2cl}
	\end{equation}  
	displays a growing correlation length diverging as $1/c$ when $T\rightarrow 0$. This is related to the fact that 
	the system orders at low temperature. The number of ground states with open boundary conditions can be shown to be equal to $2^{2L-1}$ where $L$ is the linear size of the system. In the following we shall 
	consider a very special boundary condition that simplify the computation: all the spins on the left and bottom edges of the square are fixed in the up state and the remaining ones are free. This is by no means restrictive since we are interested in bulk properties that are independent on the choice of boundary conditions. The mapping between 
	$\sigma$s and $\tau$s is particularly simple in this case: 
	\begin{equation}\label{mapping}
	\sigma_{x,y}=\prod_{l\in (x,y)} \tau_l
	\end{equation} where the product runs over all plaquettes belonging to the square with vertices $(0,0); (x,0); (y,0); (x,y)$, see Fig.1. \\
	The patch-repetition entropy in this case is
	\begin{equation}
\Sigma(\ell)=-\sum_{\mathcal{P}}  P(\mathcal{P}) \log P(\mathcal{P})\quad.
\end{equation}
where $\mathcal{P}$ is a spin configuration of a $\ell\times\ell$ square in the bulk of the system, see Fig.1. 
\begin{figure}
\onefigure[width=6.cm]{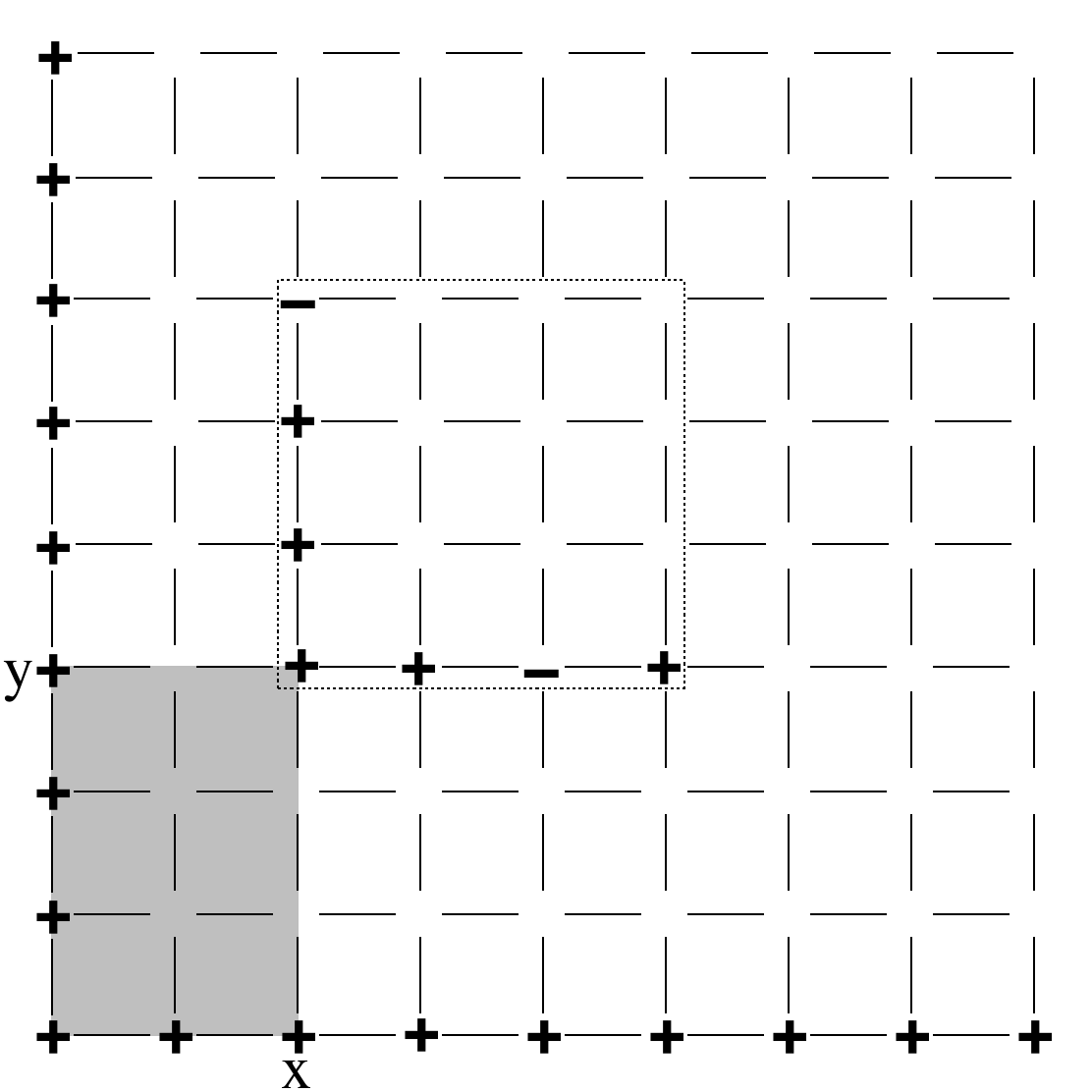}
\caption{Square plaquette lattice model. The gray region denotes all the plaquettes determining the value of 
$\sigma_{x,y}$. The pattern region corresponds to the dotted square. An example of spin configuration $\mathcal{B}$
of the dotted square is shown explicitly.}
\label{plaquette}
\end{figure}  
We shall denote $(x,y)$ its bottom left vertex.
Note that we are interested in the low temperature limit, in which the approximation of neglecting thermal fluctuations
in the definition of the PR-entropy is correct. In order to simplify the computation, we decompose the PR-entropy in 
a boundary contribution, corresponding to the configuration $\mathcal{B}$ of the spins on the bottom and left edges, and 
in a bulk contribution, corresponding to the entropy of the configuration  $\mathcal{I}$ of the remaining spins a typical given $\mathcal{B}$:
\begin{equation}
\Sigma(\ell)=\Sigma(\mathcal{B})+ \Sigma(\mathcal{I}|\mathcal{B}) \quad .
\end{equation}
It is easy to show that in the thermodynamic limit all $\mathcal{B}$s are equiprobable, {\it i.e.} 
$P(\mathcal{B})=1/2^{2\ell-1}$. Once the spins $\mathcal{B}$ are fixed, the mapping between 
the spins  $\mathcal{I}$ and the plaquette variables is one to one (just a generalization of eq. \ref{mapping}).
In consequence the PR-entropy can be computed easily:
\begin{eqnarray}
\Sigma(\ell)&=&-(\ell-1)^2\left[c\ln c+(1-c)\ln(1-c) \right]+(2\ell-1)\ln 2\nonumber\\
&\simeq &-\ell^2 c\ln c+2\ell\ln 2-\ln2 
\end{eqnarray}
where the last expression is valid at low temperature (in the last expression for each power of $\ell$ we have only retained the leading contribution in $c$). The two terms contributing to the entropy, the boundary and the bulk term, have a clear physical interpretation: the former counts the number of (equiprobable) patterns on the bottom and left edge of the square. The latter counts the number of possible patterns for a fixed boundary configurations: it corresponds to the entropy of an ideal gas of defects, corresponding to down plaquette variables. \\
Let us now investigate which lengths one can extract from the $\ell$ behavior of the PR-entropy.
From the cross-over between extensive and sub-extensive terms one obtains  
the length $\xi\simeq 2\ln2 /|c\ln c|$, see Fig.2.
\begin{figure}
\onefigure[width=8.cm]{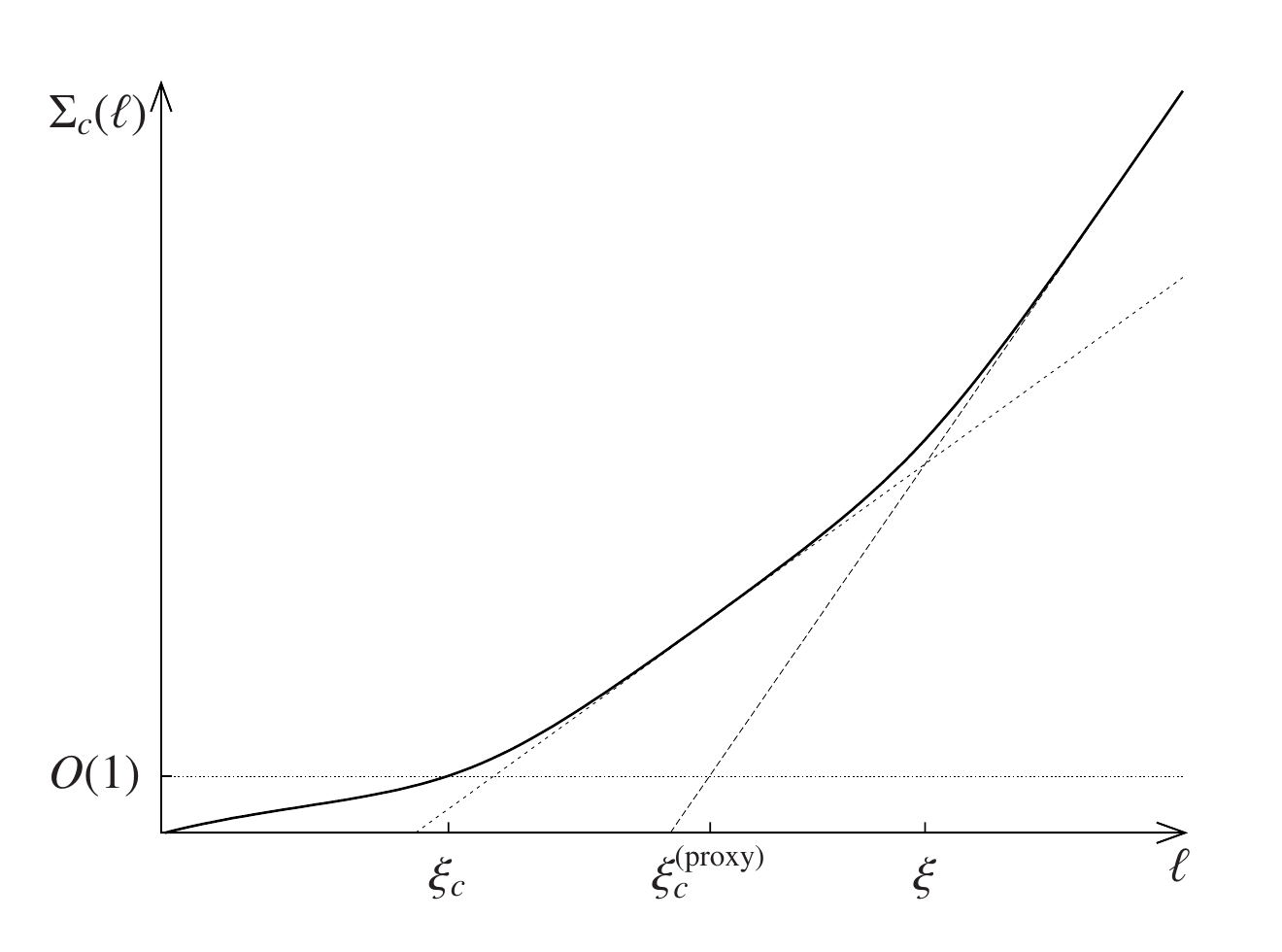}
\caption{Sketch of the log-log plot of a PR-entropy having a linear and quadratic contribution (we have also 
added the logarithmic contribution discussed in the section "patch-repetition entropy"): 
$\xi$ is the cross-over length at which $\Sigma(\ell)$ becomes extensive, $\xi_{c}^{\mbox{{\scriptsize(proxy)}}}=s_c^{-1/d}$ 
is the proxy for the cooperative length $\xi_c$. 
%***WE SHOULD NOT SHOW THE LOG BEHAVIOR
%SINCE WE DO NOT DISCUSS IT REALLY AND CHANGE THE NAME OF XI MICRO IN THE FIG.
}
\label{plaquette}
\end{figure}   
The PR-entropy increases starting from a 
value of the order of one with a finite slope, hence $\xi_c$ remains microscopic and does not grow lowering the temperature. Moreover, the length 
obtained by dimensional analysis from the extensive contribution $s_c^{-1/2}$ where $s_c=-c \ln c $ also does not play
any role, see Fig.2 for a visual summary. Thus in this model the only meaningful length that can be extracted from the PR-entropy is $\xi$.
It is interesting to compare it to the other lengths already discussed for this system \cite{Garrahan,jackgarrahan}. 
The point-to-set length, $\xi_{PS}$, was shown to increase as $1/\sqrt c$, hence much slowly than $\xi$.
Instead, the dynamical correlation scales as $1/c$, {\it i.e.} essentially as $\xi$.\\
In conclusion, the analysis of the square plaquette model shows that, at least in this case, only $\xi$ can be defined from
the $\ell$ behavior of the PR entropy. This length is much larger than  $\xi_{PS}$ and scales approximatively as 
the dynamical correlation length and the largest static correlation length, which is obtained from $G_4$.  
A similar behavior is also expected for the triangular plaquette model \cite{jackprivate}.
\section{RFOT, patch-repetition entropy and replicas}
In this section we shall derive the PR-entropy in the super-cooled regime within RFOT theory. Before doing that, we develop a 
general formalism based on the replica method to obtain $\Sigma(\ell)$. Our derivation is related to the 
Renyi complexities introduced in \cite{KL2}.\\
{\it Replica and PR-entropy}. The PR-entropy can be written using the following trick:
\begin{equation} 
\Sigma(\ell)=-\lim_{m\rightarrow 1}\frac{1}{m-1} \ln 
\left( \sum_{\mathcal P}P(\mathcal{P})^m\right)
\end{equation}
For simplicity we continue to neglect thermal fluctuations. We shall take into account their effect at the end of the derivation.
The expression inside the logarithm can be rewritten as:
\begin{eqnarray} 
&&\ln 
\left(\frac{\sum_{\mathcal P,C_1,...,C_m}e^{-\beta H({\mathcal C_1})...-\beta H({\mathcal C_m})}
\prod_{i=1}^m\Omega({\mathcal C}_i,{\mathcal P})}{\sum_{\mathcal C_1,...,C_m}e^{-\beta H({\mathcal C_1})...-\beta H({\mathcal C_m})}
}\right)\nonumber\\
&=&\ln 
\left(\frac{\sum_{\mathcal C_1,...,C_m}e^{-\beta H({\mathcal C_1})...-\beta H({\mathcal C_m})}
\Omega(\left. q_{ab}\right|_{\mathcal P}=1)}{\sum_{\mathcal C_1,...,C_m}e^{-\beta H({\mathcal C_1})...-\beta H({\mathcal C_m})}
}\right)\nonumber
\end{eqnarray}
The numerator of the above expression is the sum over $m$ replicas constrained to have an overlap equal to one 
inside the pattern region (we skipped for simplicity the sub-index denoting that the pattern is taken around a given point). Introducing $F^{(m)}(q_{ab})$, which is the Legendre transform\footnote{Actually, strictly speaking, we do not focus on the complete Legendre transform but on the average action once all short-scale degrees of freedom are integrated out, as in usual nucleation problems \cite{NPRG}.} of the free energy of the system of $m$ replicas with respect to $q_{ab}$, we find:
\begin{equation} 
\Sigma(\ell)=\lim_{m\rightarrow 1}\frac{\beta}{m-1} 
\mbox{Extr}_{q_{ab}: \left. q_{ab}\right|_{\mathcal P}=1}\left(F^{(m)}(q_{ab})-F^{(m)}(0)\right) \nonumber 
\end{equation}
where we have used that the solution of the extremization condition  in absence of constraint and
above the ideal glass transition temperature is $q_{ab}=0$. We shall now address the problem of thermal fluctuations. Taking them into account means that one 
should not count as different, patterns that are connected dynamically by short-time dynamical fluctuations. 
In order to avoid this double counting, one can lump together patterns that have an overlap larger or equal than the Edwards-Anderson parameter $q_{EA}$. The value of $q_{EA}$ can be determined from the dynamical correlation function. Of course, this procedure makes only sense when there is a clear separation of time-scales. 
The final replica expression for the PR entropy is 
therefore\footnote{In this expression we constrained all replicas to have an overlap $q_{EA}$ 
inside the pattern region. This gives the leading contribution since larger overlaps are suppressed
exponentially (in the size of the pattern region).}:
\begin{equation} \label{variational}
\Sigma(\ell)=\beta\frac{d}{dm} 
\left.
\hspace{-0.1cm} \mbox{Extr}_{q_{ab}: \left. q_{ab}\right|_{\mathcal P}=q_{EA}}\hspace{-0.15cm}\left(F^{(m)}(q_{ab})-F^{(m)}(0)\right)\right|_{m=1}
\end{equation}
{\it PR entropy and RFOT theory}. In order to perform the computation of $\Sigma(\ell)$ we need $\beta F^{(m)}( q_{ab})$. This is not known in general since its computation would require to solve the full theory by integrating over almost all fluctuations. However, in the Kac-limit \cite{KacSilvio} or on general grounds, one can  
argue that $\beta\Delta F=\beta \left( F^{(m)}(q_{ab})- F^{(m)}(0)\right)$ should have a form like: 
\begin{equation}
\label{eq_GLfunctional}
\beta\Delta F=\int \frac{d^d x}{a_0^d} \bigg\{\frac{a_0^2}{2} \sum_{a,b=1}^m  \left( \partial q_{ab}(x)\right)^2 +  \tilde V(q_{ab}(x)) \bigg\}
\end{equation}
The first term represents the tendency of replica to remain coupled. We used a squared gradient to mimic this effect. Although this is just an approximation, other forms, e.g. non-local but still elastic ones, would lead to the same result as long as they have a finite range $a_0$, which physically corresponds to the characteristic microscopic length, {\it i.e.} the length corresponding to the first peak in the radial distribution function (henceforth
we shall put $a_0=1$ and measure lengths in unit of $a_0$).
Instead, in the case of a more general elastic term leading to an interface cost scaling with an exponent $\theta<d-1$ the results below are modified: $d-1$ has to be replaced by $\theta$ in the final
expression for $\Sigma(\ell)$. The precise expression of the second term, the potential, 
actually is not important. The only thing which matters is that  in the $m\rightarrow 1$ limit and within the hypothesis of replica symmetry, 
$q_{ab}(x)=q(x)$ $\forall a\neq b$, the potential $\tilde V(q_{ab}(x))$ simplifies to 
$(m-1)V(q(x))$ with $V(q)$ having the shape shown in Fig.3. 
\begin{figure}
\onefigure[width=8.cm]{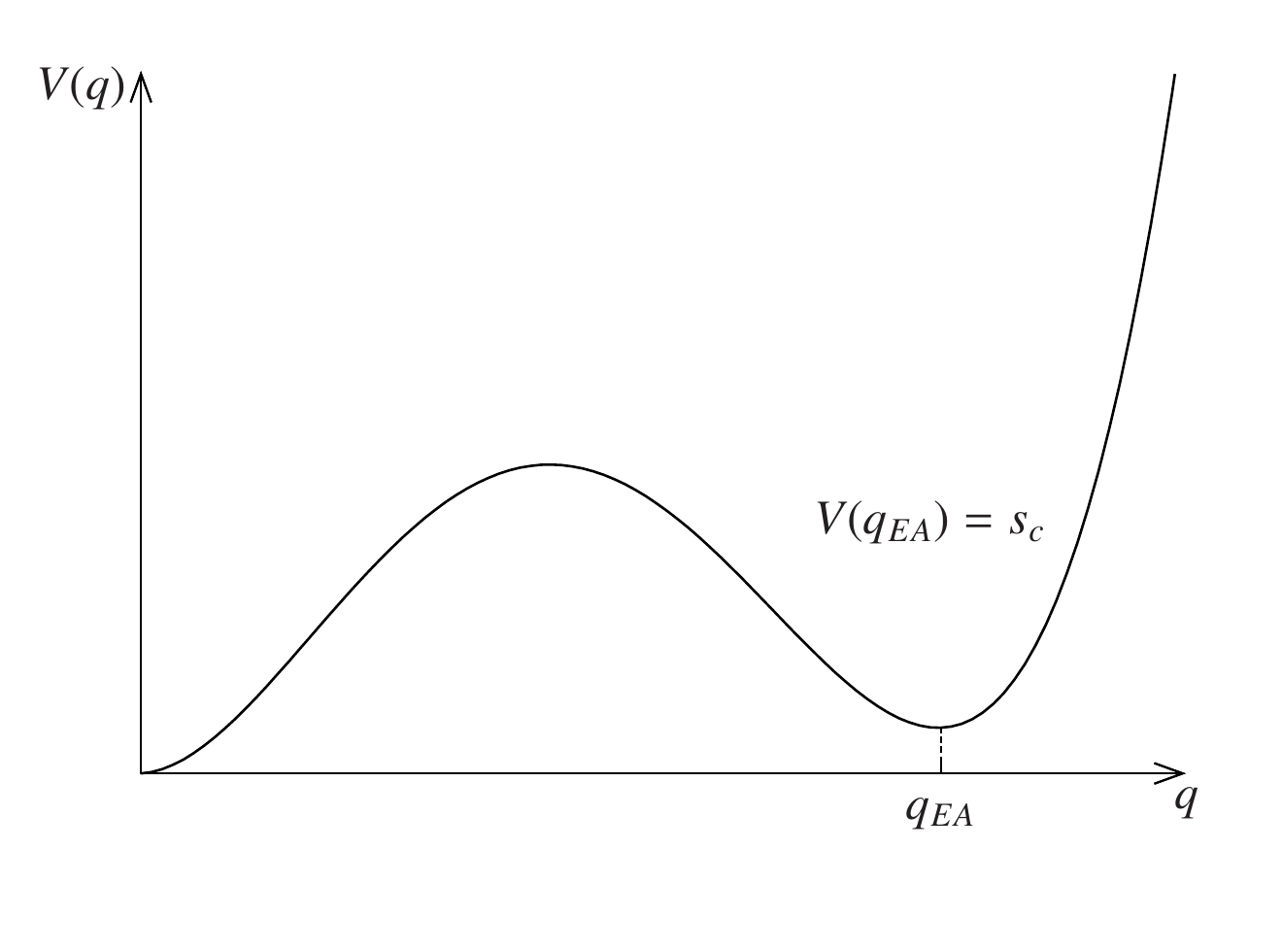}
\caption{Sketch of the shape of $V(q)$.
%***CUT THE T-TK IN THE FIG. PUT ARROWS ON THE AXIS AND REPLACE
%Q* with QEA
}
\label{plaquette}
\end{figure}   
This is indeed what happens in the Kac-limit investigated by Franz \cite{KacSilvio} and it is what one would expect  if RFOT theory retains a validity in finite dimensions.  Previous studies have shown that the value of $q$ at the local secondary minimum of $V(q)$ corresponds to $q_{EA}$ and that $V(q_{EA})$ is the intensive value of the configurational entropy $s_c$, which goes to zero linearly at the ideal glass transition temperature $T_K$ \cite{Monasson,FranzParisi}. \\
By assuming that replica symmetry is not broken and considering, without loss of generality, a spherical pattern region we find that the solution of the variational problem (\ref{variational}) satisfies the equation:
\begin{equation}
-\frac{d^2q}{dr^2} -\frac{d-1}{r}\frac{dq}{dr}+ V'(q)=0
\end{equation}
Close to $T_K$ the local minimum is almost degenerate with the global one and $q$ 
is found to vary on large length-scales. Thus, 
making the thin-wall approximation \cite{Coleman}, {\it i.e. } by dropping the first derivative, one 
obtains a one dimensional Newton equation for a particle moving in the potential $- V(q)$ during an effective time $r$. For the optimal solution $ V(q)$ must vanish for $r\rightarrow \infty$. Thus, the solution $q(r)$ corresponds to the trajectory of a particle that starts at $q=q_{EA}$ at $r=0$, with a 
"kinetic energy" equal to $ V(q_{EA})$, and reaches $q=0$ at infinite time with vanishing kinetic energy. 
Using this result and following standard manipulations \cite{Coleman} one finds (in three dimension):
$$
\beta F^{(m)}(q_{ab})-\beta F^{(m)}(0)=(m-1)\left(\frac 4 3 \pi \ell^3  s_c+4\pi \ell^2 Y\right)
$$
where $Y=\int_0^{q_{EA}} \sqrt{2 V(q)}dq$. Using eq. (\ref{variational}) we finally obtain the RFOT prediction for the
PR-entropy:
\begin{equation}\label {PRentropy}
\Sigma(\ell)=\frac 4 3 \pi \ell^3  s_c+4\pi \ell^2 Y
\end{equation}
A part from irrelevant numerical constants, the $\ell$ dependence of $\Sigma(\ell)$ is the three dimensional analog of the one found for the square plaquette model in the previous section. In consequence, as previously, the only meaningful length that one can extract from the PR-entropy is $\xi=3Y/s_c$. Remarkably, this scaling with $s_c$, which is valid in any dimension, is the same one of the point-to set length  $\xi_{PS}$ (even in the case $\theta<d-1$). Thus, within RFOT, these two length-scale coincide. \\
{\it The fuzzy mosaic state}.
The description of the super-cooled liquid state resulting from RFOT is often 
represented or envisioned as a mosaic of patches, each one localized (temporarily) in a given
amorphous configuration\cite{RFOTreviewWolynes,RFOTreviewBB}. Showing whether such a state does exist and what is its real space structure remains a pressing open problem in the field. The PR-entropy should be 
a good probe to verify the mosaic representation. 
Unexpectedly, the explicit computation of $\Sigma(\ell)$ within RFOT
leads to an expression which is not compatible with such real space structure. 
In order to clarify this point, let us obtain our result again, but in a different way by 
rewriting $\Sigma(\ell)$ as the sum of the 
PR entropy for the boundary, $\Sigma_B(\ell)$, and the PR entropy of the interior of the sphere given a fixed typical boundary, $\Sigma_{C|B}(\ell)$. For a system with short range 
interactions once the boundary of a closed region is fixed any information with the exterior is lost, thus fixing a boundary 
is like fixing all the exterior. Hence, $\Sigma_{C|B}(\ell)$ coincides with the configurational entropy of the particles inside a cavity when all particles
outside are blocked in a typical equilibrium configuration. This, actually, is the protocol used to compute point-to-set lengths.   
The analysis of \cite{FranzMontanari} shows that close to the ideal glass transition  $\Sigma_{C|B}(\ell)$  is zero for  $\ell<\xi_{PS}$ and equal to $\frac 4 3 \pi \ell^3  s_c-4\pi \ell^2 Y$ for  $\ell>\xi_{PS}$. This is exactly what one expects from the definition of the point-to-set length: boundary conditions determines the bulk amorphous configuration below, but not above, $\xi_{PS}$. 
The behavior of  $\Sigma_B(\ell)$ can be obtained by generalizing our previous computation: one has to solve a variational
problem in which the profile $q(x)$ is constrained to be equal to $q_{EA}$ in a small annular region. The result is 
that $\Sigma_B(\ell)$ is equal to $\frac 4 3 \pi \ell^3  s_c+4\pi \ell^2 Y$ below the point-to set length and
equal to  $8\pi \ell^2 Y$ above it. Summing the two contributions, one finds back (\ref{PRentropy}), as it has to be. 
Remarkably, the single contributions display a transition at the point-to-set length but their sum, the PR entropy, has no
singularity but just a cross-over.
The behavior of $\Sigma_B(\ell)$ is quite surprising. Close to $T_K$, for $\ell=\xi_{PS}$, 
the number of typical boundary patterns is very large. Actually, it starts to be much 
larger than one as soon as $\ell>1$.  This suggests that given a snapshot (even one coarse grained in time, 
as discussed in \cite{KL2}) no clear boundaries between local amorphous states do exist\footnote{The non-existence
of boundaries was already proposed by S. Franz \cite{franzguilhem} on the basis of the result that the average energy (or of any other local observable) inside a cavity with amorphous boundary conditions coincides with the average energy for the free system. 
%At first sight, no boundaries can exist if one does not find any change in the average energy (or in any other local observable) by changing the size of the cavity, in particular across $\xi_{PS}$. However,  
%if 
However, by taking into account that for $\ell<\xi_{PS}$ sometimes, even though rarely,
boundaries are present inside the cavity, one finds that this result is not really in contradiction {\it per se} with the existence of boundaries.}. Were they present, one would expect $\Sigma_B(\ell)$ to be related to the number of possible 
different amorphous states on scale $\ell$ for $\ell\ll\xi_{PS}$. What is this number ${\mathcal N}_S(\ell)$? 
Naively, one could think that there is a one to one correspondence between the tiles of the mosaic (amorphous states on scale $\xi_{PS}$) and the possible states on scale $\ell$. 
This would lead to  $\log {\mathcal N}_S(\ell)=s_c \xi_{PS}^d$, which clearly does not hold. 
A more refined reasoning takes into account that different tiles can contain the same patterns on length-scales smaller than $\xi_{PS}$. 
The simplest expectation would then be that  ${\mathcal N}_S(\ell)=({\mathcal N}_S(\xi_{PS}))^{(\ell/\xi_{PS})^d}=e^{s_c\ell^d}$ \cite{KL,KL2}. However, this also does not hold. The only way to get our result for $1\ll \ell \ll \xi_{PS}$ is that  ${\mathcal N}_S(\ell)\simeq({\mathcal N}_S(\xi_{PS}))^{\frac 1 2 (\ell/\xi_{PS})^{d-1}}=e^{4 \pi Y\ell^{d-1}}$, a relation 
that seems hard to justify starting from a picture in terms of a 
mosaic with well-defined boundaries.\\
%Although this assumption cannot be excluded, it is certainly not very plausible.\\ 
The absence of boundaries between local amorphous states can be backed up by an independent
argument based on RFOT and our recent results on random pinning glass transitions\cite{BC}. We have recently shown that 
within RFOT by pinning at random particles of an equilibrated configuration one can induce ideal glass transitions for the
remaining free particles at temperatures higher than $T_K$. Now, imagine that one is able to show the existence of  boundaries for the unconstrained liquid. The only way to do that consists in finding local fluctuations of some sort (of energy, entropy, density, etc.) which pinpoint the presence 
of the boundary in real space. A paradox arises when considering the effect of random pinning. 
By increasing the fraction of pinned particles one can approach (at fixed temperature) the glass transition arbitrarily close and, correspondingly, 
obtain an arbitrarily large point-to-set length and very large boundaries. The problem is that this has to be true
also for the initial configuration, since this is an equilibrated configuration for the free particles,
independently on the fraction of pinned ones.  Thus, the initial configuration must contain boundaries on all 
length-scales. This seems---and actually is---impossible. It would imply that 
above $T_K$ there exists infinite range spatial correlations and, consequently, that  the variance of fluctuations on 
length-scales $\ell$ of the observable correlated with the presence of boundaries 
never display the scaling $\ell^{d/2}$. An awkward result, especially if one takes into account
fluctuation-dissipation relations: a scaling different from $\ell^{d/2}$ for fluctuations 
implies divergent response functions. For example, if the local observable was the energy then 
one would obtain an infinite specific heat above $T_K$. \\
We conclude that boundaries between amorphous states do not exist: the mosaic structure is very fuzzy\footnote{
The RFOT version corresponding to $\theta=d/2$ also assumes no clear boundaries, or alternatively, boundaries 
everywhere and on all scales below $\xi_{PS}$ \cite{RFOTreviewWolynes}.}.  
If one wants to illustrate the real-space structure resulting from RFOT with a metaphor one 
should think to a drawing where colors change rapidly and in an apparently random fashion instead 
than to a mosaic of patches with different colors---it would resemble more to 
a Pollock's abstract painting than to a Mondrian's one.  
The presence of amorphous order in this drawing would be unveiled focusing on all regions where 
the same (apparently random) boundary appears: one would typically find the same pattern 
inside the boundary if this has a linear size smaller than $\xi_{PS}$.
\section{Conclusion}
We studied the behavior of the PR-entropy in the square plaquette model and within RFOT theory. 
In both cases we find that the $\ell$ dependence of $\Sigma(\ell)$ is given by 
an extensive term followed by a sub-leading one
scaling as the area of the pattern region. We found that the only meaningful length one can define, $\xi$, is 
the one corresponding to the cross-over between leading and sub-leading behavior in $\ell$ of $\Sigma(\ell)$. 
This length, however, is neither $\xi_c$, at which the PR-entropy becomes of order one
(this remains featureless, microscopic and does not grow), nor its proxy $s_c^{-1/d}$. 
Comparing $\xi$ to the other lengths involved, we discovered that it
coincides with the point-to-set length in the RFOT case only and it always corresponds to 
the largest static length, a fact that is possibly valid in general.
Our results provide new insights, which are 
particularly relevant for RFOT theory, on the possible real space structure of super-cooled liquids. 
They clarify that its illustration in terms of a mosaic of different patterns is not qualitative correct. 
For lack of a better name we dub the correct one, holding at least within RFOT, the fuzzy mosaic state.  
 
 %\begin{figure}
%\onefigure{epl-template.eps}
%\caption{Figure caption.}
%\label{fig.1}
%\end{figure}

\acknowledgments
We thank J.-P. Bouchaud, R. Jack, J. Kurchan and P. Wolynes for helpful discussions. 
GB acknowledges support from ERC grant NPRG-GLASS.

\end{document}